\documentclass[a4paper]{amsart}
\usepackage[T1]{fontenc}
\usepackage[latin1]{inputenc}
\usepackage{mathrsfs,dsfont,amsthm}
\theoremstyle{definition}
\newtheorem{definition}{Definition}
\theoremstyle{plain}
\newtheorem{proposition}{Proposition}
\theoremstyle{plain}
\newtheorem{theorem}{Theorem}
\theoremstyle{plain}
\newtheorem{lemma}{Lemma}
\theoremstyle{remark}
\newtheorem{remark}{Remark}
\theoremstyle{plain}
\newtheorem{acknowledgement}{Aknowledgement}
\newenvironment{bew}[2]{\removelastskip\vspace{6pt}\noindent
 {\it Proof  #1.}~\rm#2}{\par\vspace{6pt}}
\newlength{\enumskip}
\begin{document}

\title{{Localization of quantum wave packets}}

\author{Eric Str\"{a}ng}
\address{Institut f\"{u}r Theoretische Physik, Universit\"{a}t Ulm, 
Albert-Einstein-Allee 11, D89069-Ulm, Germany}
\email{eric.straeng@uni-ulm.de}

\begin{abstract}
We study the semiclassical propagation of squeezed Gau{\ss}ian states.
We do so by considering the propagation theorem introduced by Combescure
and Robert \cite{CR97} approximating the evolution generated by the
Weyl-quantization of symbols $H$. We examine the
particular case when the Hessian $H^{\prime\prime}\left(X_{t}\right)$
evaluated at the corresponding solution $X_{t}$ of Hamilton's equations of 
motion is periodic in time. Under this assumption,
we show that the width of the wave packet can remain small up
to the Ehrenfest time. We also determine conditions for ``classical
revivals'' in that case. More generally, we may define recurrences 
of the initial width. Some of these results include the case of unbounded 
classical motion. In the classically unstable case we recover an exponential
spreading of the wave packet as in \cite{CR97}. 
\end{abstract}
\maketitle

\section{Introduction}

Localization of quantum states in phase space is a prerequisite
in some semiclassical treatments of quantum evolution. In the classically chaotic case,
the width of an initially localized Gau{\ss}ian increases exponentially
\cite{CR97} up to the so called Ehrenfest time $T_{E}$, i.e., the time up to 
which quantum dynamics can be approximated by classical dynamics.  In the case
of regular classical motion it can be shown that this width grows algebraically
in the semiclassical parameter $\hbar$ up to $T_{E}$.

In certain applications, it is necessary to propagate states semiclassically for long times. 
This demands control of the width of the state. An example is the construction of quasi-modes 
proposed by Paul and Uribe \cite{PaUr93}. One refers to a quasi-mode as a state $\psi$ 
which is a solution of a corresponding spectral problem of an operator $\widehat{H}$ up to some small discrepancy $\delta$, i.e., 
\begin{eqnarray*}
 \left\Vert\widehat{H}\psi-E\psi\right\Vert & < & \delta 
\end{eqnarray*} which insures, in the case of discrete spectrum, that there exist at least one eigenvalue of $\widehat{H}$ 
in the interval $\left[E-\delta,\,E+\delta\right]$. 
The approximation proposed by Paul and Uribe uses the semiclassical propagation of coherent 
states over closed classical trajectories leading to the well known Bohr-Sommerfeld quantization
rule in one dimension. The eigenvalues $E_{n}$ of the Hamiltonian $\widehat{H}$ are given 
by the quantization condition
\begin{eqnarray}
\int_{\mathscr{C}_{E_{n}}}p\,\mathrm{d}q & = & 2\pi\hbar\left(n+\frac{1}{2}\right)+\mathcal{O}\left(\hbar^{2}\right),\quad n\in\mathds{N},
\end{eqnarray}
where the energy shells $\mathscr{C}_{E}=\left\lbrace \left(p,\,q\right):\,H\left(p,\, q\right)=E\right\rbrace $ are closed curves.

Certain examples suggest that there exist systems (other than the harmonic oscillator)
where the propagated width of an initial Gau{\ss}ian remains small for long times. 
Such a behavior is exhibited by the propagation of Gau{\ss}ians generated by some 
perturbed periodic Schr\"{o}dinger operators like the Wannier-Stark Hamiltonian
\begin{eqnarray*}
\widehat{H}_{\mathrm{WS}}\left( \varepsilon \right) & = & -\frac{\hbar^{2}}{2}\Delta_{x}+V_{\Gamma}\left( x\right)+\varepsilon x 
\end{eqnarray*}
in the limit of small perturbations $\varepsilon$. Here $V_{\Gamma}\left(x \right)$ denotes a 
periodic potential with respect to a lattice $\Gamma\cong \mathds{Z}^{d}$. It is known that
the band structure of the spectrum of the unperturbed operator is preserved for small enough 
perturbations $\varepsilon$. Numerical studies \cite{Har04,Witt04} show that an initially
localized Gau{\ss}ian in momentum space defined on such an energy band apparently remains 
Gau{\ss}ian for long times. The evolving states carry out oscillations. The center of the 
Gau{\ss}ian can oscillate in position space describing so called Bloch oscillations \cite{Blo28}. 
Alternatively, the width of the Gaussian can oscillate, whereupon the center remains fixed, 
describing so called breathing modes. In both cases, the state returns to the initial 
state after an oscillation period up to a small error.

We study the evolution generated by a class of operators on 
$L^{2}\left(\mathds{R}^{d}\right)$
defined as Weyl quantizations of classical symbols $H\left(X\right)$
with the property\begin{eqnarray}
H^{\prime\prime}\left(X_{t}\right) & = & 
H^{\prime\prime}\left(X_{t+T}\right)
,\quad\forall t\in\mathds{R},\label{eq:percond}
\end{eqnarray}
where $X_{t}$ denotes the solutions to Hamilton's equations of motion and 
$H^{\prime\prime}\left(X\right)$ is the Hessian of $H$ with respect to $X$.
In one dimension, the condition (\ref{eq:percond}) is satisfied by bounded 
classical motion or unbounded
motion in a periodic potential. We particularly study the evolution
of initial Gau{\ss}ian (or squeezed) states semi-classically, i.e.,
asymptotically as $\hbar\searrow0$, when $t\nearrow\infty$.
We focus our attention to the spreading of such wave packets.

In sections \ref{sec:Preliminaries:-semiclassical-propagation} and
\ref{sec:A semiclassical propagation theorem} we shortly
review the semiclassical propagation of Gau{\ss}ian coherent states. In
section \ref{EhrT} we give some known results on the validity of the
approximation. We then make, in section \ref{sec:Floquet-theory} and 
section \ref{sec:A-Uniform-bound}, statements about the approximate 
Gau{\ss}ian state given by this semiclassical propagation using Floquet 
theory. These properties are then brought back to the true evolution 
in section \ref{sec:Discussion}.

\subsection{\label{sec:Preliminaries:-semiclassical-propagation}Preliminaries:
semiclassical propagation of wave packets}

We will work in the context of self adjoint operators defined on $L^{2}\left(\mathds{R}^{d}\right)$
that are $\hbar-$Weyl quantizations of symbols. To a  smooth 
$\left(C^{\infty}\right)$ classical
symbol $b\left( X\right)$, i.e., a function on the phase space
$T^{*}\mathds{R}^{d}\cong\mathds{R}^{2d}$, there corresponds an operator
on $L^{2}\left(\mathds{R}^{d}\right)$, $\widehat{b}:=\mathrm{Op}_{\hbar}^{w}\left[b\right]$,
defined by\begin{eqnarray*}
\mathrm{Op}_{\hbar}^{w}\left[b\right]\psi\left(x\right) & := & \frac{1}{\left(2\pi\hbar\right)^{d}}\int_{\mathds{R}^{2d}}b\left(\frac{x+y}{2},\,\xi\right)\psi\left(y\right)e^{\frac{\mathrm{i}}{\hbar}\left(x-y\right)\xi}\,\mathrm{d}y\mathrm{d}\xi.\end{eqnarray*}

The following conditions will be assumed \label{conditions}:

\begin{enumerate}
\item \vspace{\enumskip}The classical Hamiltonian $H:\,\mathds{R}^{2d}\rightarrow\mathds{R}$ is a smooth function.
\item \vspace{\enumskip}$H\in S\left(m\right)$, i.e., for all multi-indices
$\alpha$, $\beta$, there exists $K_{\alpha ,\,\beta}>0$ such that \[
\left|\partial_{p}^{\alpha}\partial_{q}^{\beta}H\left(p,\, q\right)\right|\leq K_{\alpha ,\,\beta}\left(1+\left|p\right|^{2}+\left|q\right|^{2}\right)^{\frac{m}{2}}. \]

\item \vspace{\enumskip}The corresponding classical equation of motion is given
by \begin{eqnarray*}
\frac{\mathrm{d}X_{t}}{\mathrm{d}t} & = & \mathcal{J}H^{\prime}\left(X_{t}\right),
\end{eqnarray*}
where $\mathcal{J}$ is the symplectic
unity \begin{eqnarray*}
\mathcal{J} & := & \left(\begin{array}{cc}
0 & -\mathds{1}_{d\times d}\\
\mathds{1}_{d\times d} & 0\end{array}\right),
\end{eqnarray*}
 and $H^{\prime}$ is the gradient of $H$ with respect to $X$. Furthermore, 
we denote by $\Phi_{H}^{t}:\,\mathds{R}^{2d}\rightarrow\mathds{R}^{2d},
\,X_{0}\mapsto X_{t}=\Phi_{H}^{t}\left(X_{0}\right)$ the corresponding classical flow.
\item \vspace{\enumskip}The $\hbar-$Weyl quantization of $H\left(X\right)$,
$\widehat{H}:=\mathrm{Op}_{\hbar}^{w}\left[H\right]$, is an essentially self-adjoint
 operator on $L^{2}\left(\mathds{R}^{d}\right)$
and generates a unitary time evolution $\forall t\in\mathds{R},$\begin{eqnarray*}
\widehat{U}\left(t \right) & : & L^{2}\left(\mathds{R}^{d}\right)\rightarrow L^{2}\left(\mathds{R}^{d}\right)\\
 &  & \psi\left(0\right)\mapsto\psi\left(t\right)\end{eqnarray*}
corresponding to the Schr\"{o}dinger equation\begin{eqnarray}
\mathrm{i}\hbar\frac{\partial\psi}{\partial t} & = & \widehat{H}\psi,\label{eq:Schroedinger}\end{eqnarray}
i.e., we will write $\widehat{U}\left(t\right)=e^{-\frac{\mathrm{i}}{\hbar}\widehat{H}t}.$\vspace{\enumskip}
\end{enumerate}

The Weyl calculus also allows a representation of quantum mechanical wave
functions on phase space. This is given by the Wigner function of
the state $u\in L^{2}\left(\mathds{R}^{d}\right),$
\begin{eqnarray}
W\left[u\right]\left(p,\, q\right) & := & \int_{\mathds{R}^{d}}\overline{u}\left(q+\frac{y}{2}\right)u\left(q-\frac{y}{2}\right)e^{\frac{\mathrm{i}}{\hbar} p y}\,\mathrm{d}y\label{WigRep}.\end{eqnarray}
The Wigner function of a Gau{\ss}ian defines a positive measure on
phase space.

\subsection{\label{sec:A semiclassical propagation theorem}A semiclassical propagation theorem}

With these assumptions we give a short summary of the method
of semiclassical propagation of coherent states introduced by Combescure
and Robert \cite{CR97}. Similar constructions have also been considered
in the past. See, e.g., Hagedorn \cite{Ha81,Ha85} and references
therein.

The idea is to expand the exact Hamiltonian $\widehat{H}$ along the
classical flow generated by the symbol $H$ up to second order, whereupon
the approximate time dependent Hamiltonian, $\widehat{H}_{2}\left(t\right)$,
is the $\hbar-$Weyl quantization of
\begin{eqnarray*}
H_{2}\left(t,\,Y\right) & := & H\left(X_{t}\right)+\left(Y-X_{t}\right)^{T}
H^{\prime}\left(X_{t}\right)+\frac{1}{2}\left(Y-X_{t}\right)^{T}H^{\prime\prime}
\left(X_{t}\right)\left(Y-X_{t}\right).
\end{eqnarray*}
The propagation  of normalized wave functions (squeezed states)%
\footnote{We will, with some lack of rigor, call these states Gau{\ss}ian,
coherent or squeezed without distinction.%
} of the form 
\begin{eqnarray}
\psi_{Z}^{\left(p,\, q\right)}\left(x\right) & := & 
\frac{\det\left(\Im\left(Z\right)\right)^{\frac{1}{4}}}
{\left(\pi\hbar\right)^{\frac{d}{4}}}e^{\frac{\mathrm{i}}{\hbar}
\left(p^{T}\left(x-q\right)+\left(x-q\right)^{T}\frac{Z}{2}\left(x-q\right)\right)},
\,Z\in\Sigma_{d},\,\left(\begin{array}{c}p \\ q\end{array}\right)\in\mathds{R}^{2d}
\label{eq:squstate}\end{eqnarray}
by quadratic Hamiltonians is well known \cite{Fol89}. By $\Sigma_{d}$, 
we mean the $d-$dimensional Siegel upper half space, i.e., the set of symmetric
$d\times d$ matrices with positive, non-degenerate imaginary
part \cite{Fol89}. The quadratic form $Z$
describes the shape of the wave packet and it should be underlined
that it is independent of $\hbar$. 
In its Wigner representation (see eq.(\ref{WigRep}))
$\psi_{Z}^{\left(p,\, q\right)}$
is a Gau{\ss}ian centered around $X=\left(\begin{array}{c}
p\\
q\end{array}\right).$ This Wigner function is given by
\begin{eqnarray*}
W\left[\psi_{Z}^{X}\right]\left(Y\right) & = & \left(\frac{1}
{\pi\hbar}\right)^{d}\exp\left(-\frac{1}{\hbar}
\left(Y-X\right)^{T}G\left(Y-X\right)\right)
\end{eqnarray*}
where
\begin{eqnarray}
G & := & \left(\begin{array}{ccc}
\Im\left(Z\right)^{-1} & \quad & -\Im\left(Z\right)^{-1}\Re \left(Z\right)\\
\quad & \quad & \quad\\
-\Re\left(Z \right)\Im\left(Z\right)^{-1} & \quad &
\Im\left(Z \right)+\Re\left(Z \right)\Im
\left(Z \right)^{-1}\Re\left(Z \right) \end{array}\right)
\end{eqnarray}
is independent of $\hbar$.

The unitary evolution, $\widehat{U}_{2}\left(t\right)$, generated
by $\widehat{H}_{2}\left(t\right)$, acts on a squeezed state by translation and
metaplectic action, i.e.,
\begin{eqnarray}
\widehat{U}_{2}\left(t\right) & = & e^{\frac{\mathrm{i}}{\hbar}\Theta\left(t\right)}
\widehat{\mathcal{T}}\left( X_{t}\right)
\widehat{\mathcal{M}}\left(S_{t}\right)
\label{eq:metaplect}\end{eqnarray}
where $\widehat{\mathcal{T}}$ is the translation operator on $\mathds{R}^{2d}$, i.e., $\forall Y \in\mathds{R}^{2d},\,
=:\left(
\begin{array}{c}
\xi \\
q
\end{array}
\right)$ and $u\in L^{2}\left(\mathds{R}^{d}\right)$ we have
\begin{eqnarray*}
 \widehat{\mathcal{T}}\left(Y\right)W\left[u\right]\left(Z\right)
 & := & e^{\frac{\mathrm{i}}{\hbar}\left(\xi^{T} \widehat{x}-q^{T}\widehat{p}\right)}W\left[u\right]\left(Z\right)\\
 & = & W\left[u\right]\left(Z-Y\right),
\end{eqnarray*} 
where we denote by $\widehat{p}$ the momentum operator and by $\widehat{x}$ the position operator. The metaplectic
operator $\widehat{\mathcal{M}}\left( F\right)$ is the quantization of a linear symplectomorphism on $\mathds{R}^{2d}$ given by the symplectic matrix $F$. These operators form a double-valued unitary representation\footnote{A thorough description of the action of the evolution generated by quadratic operators on Gau{\ss}ians and the metaplectic representation can be found in \cite{Fol89}.} of the linear symplectomorphism of $\mathds{R}^{2d}$. $S_{t}$ denotes the flow differential. The classical flow $\Phi_{H}^{t}$ is a symplectomorphism which ensures that the flow differential is a symplectic matrix. Furthermore, $S_{t}$ satisfies 
\begin{eqnarray}
\frac{\mathrm{d}S_{t}}{\mathrm{d}t} & = & \mathcal{J}H^{\prime\prime}
\left(X_{t}\right)S_{t},\label{eq:action}\\
\,S_{0} & = & \mathds{1}_{2d\times2d}.\nonumber 
\end{eqnarray} In the prefactor of eq. (\ref{eq:metaplect}) we have used
\begin{eqnarray*}
\Theta\left(t\right)&:=&\mathcal{W}\left(t\right)+\hbar\mu, 
\end{eqnarray*}
where  
\begin{eqnarray*}
\mathcal{W}\left(t\right) & := & \int_{0}^{t} \left(p_{\tau}^{T}\dot{q}_{\tau}-H\left(p_{\tau},\, 
q_{\tau}\right)\right)\, d\tau,
\end{eqnarray*}
is the action of the classical trajectory and $\mu$ is the Maslov
index of the classical trajectory. We have here expressed the solution 
of Hamilton's equations in terms of the canonical variables
\begin{eqnarray*}
\left(\begin{array}{c}
p_{t}\\
q_{t}\end{array}\right) & := & X_{t}.
\end{eqnarray*}

Acting with (\ref{eq:metaplect}) on (\ref{eq:squstate}), one obtains at time $t$ a new Gau{\ss}ian state
$e^{\frac{\mathrm{i}}{\hbar}\Theta\left(t\right)}\psi_{Z_{t}}^{X_{t}}$ up to a 
phase. The Gau{\ss}ian is centered around $X_{t}$ in phase space 
and has a quadratic form $Z_{t}\in\Sigma_{d}$, is given explicitly by the group action 
\cite{Fol89}

\begin{eqnarray*}
Z_{t} & = & S_{t}\left[Z_{0}\right]\\
 & = & \left(A_{t}Z_{0}+B_{t}\right)\left(C_{t}Z_{0}+D_{t}\right)^{-1}\quad
,\,Z_{0}\in\Sigma_{d},
\end{eqnarray*}
i.e., linear fractional transformation on $\Sigma_{d}$. The matrix
$S_{t}$ is here written by means of the $d\times d$ blocks $A_{t}$, $B_{t}$, $C_{t}$ and $D_{t}$, i.e.,
\begin{eqnarray}
S_{t} & = & \left(\begin{array}{cc}
A_{t} & B_{t}\\
C_{t} & D_{t}\end{array}\right).
\end{eqnarray}
The quadratic form of the Wigner transform of $\psi_{Z_{t}}^{X_{t}}$, is explicitly given by
\begin{eqnarray}
G_{t} & = & \left(S_{t}^{-1}\right)^{T}G_{0}S_{t}^{-1}\label{eq:width}.
\end{eqnarray}

The evolution of the approximate wave packet can be perceived as being 
generated by rotation and scaling of the Gau{\ss}ian profile in 
$\mathds{R}^{2d}$.

The difficulty of this scheme resides in the control of errors made 
by the approximation of $\widehat{U}\left(t\right)$ in terms of
 $\widehat{U}_{2}\left(t\right)$. We do not dwell on 
the details, but just state the result (for proof and a thorough description see 
\cite{CR97}). The approximation can be checked to all orders in 
$\hbar$. For this purpose, one defines the following approximating state
\begin{eqnarray*}
\Psi^{\left(N\right)}\left(t,\, x\right) & := & e^{\frac{\mathrm{i}}{\hbar}\Theta\left(t\right)}\sum_{0\leq j\leq N}\hbar^{\frac{j}{2}}\pi_{j}\left(X_{t},\,\frac{x}{\sqrt{\hbar}},\, t\right)\psi_{Z_{t}}^{X_{t}}\left(x\right),
\end{eqnarray*}
where $\pi_{j}\left(X_{t},\,\frac{x}{\sqrt{\hbar}},\, t\right)$
are polynomials in $x\left/\sqrt{\hbar}\right.$ and $X_{t}$
of degree smaller than or equal to $3j$ with time dependent coefficients. 

\begin{theorem}
\label{thm:propagation_control_CR}(Combescure, Robert) Under the above
mentioned assumptions and for an initial Gau{\ss}ian state $\psi_{Z_{0}}^{X_{0}}$
centered in the phase space representation at $X_{0}\in\mathds{R}^{2d}$, for every 
$N\in\mathds{N}$ there exists $C<\infty$ such that $\forall\hbar\in\left(0,\,\hbar_{0}\right],\,\hbar_{0}>0$,
\begin{eqnarray}
\left\Vert \widehat{U}\left(t\right)\psi_{Z_{0}}^{X_{0}}-\Psi^{\left(N\right)}\left(t\right)\right\Vert  & 
\leq & C_{N}\hbar^{\frac{N+1}{2}}te^{3\gamma t}\label{eq:normunitaries}
\end{eqnarray}
where $0\leq\gamma<\infty$ is the Lyapunov exponent of the classical motion. 
\end{theorem}

We recall that a Lyapunov exponent is a measure of the exponential stability of the solutions of a 
differential equation upon change of initial conditions. In the case of classical motion, this is given by the Lyapunov
exponent defined as
\begin{eqnarray}
\gamma & := & \max_{k}\left[  \lim_{t\rightarrow\infty}\sup\left(  \frac{\ln\left(s_{k}\left( t\right)\right)}{t}\right) \right] \label{defclasslyapunov}
\end{eqnarray}
where $s_{k}\left( t\right)$ are the singular values of $S_{t}$. The Lyapunov exponent $\gamma$ hence satisfies
\begin{eqnarray}
 \left\Vert S_{t}\right\Vert_{\mathrm{HS}} & < & c_{0}e^{\gamma \left\vert t\right\vert}\label{lyapunovinequality}
\end{eqnarray}
where $c_{0}<\infty$ is a positive constant. We denote by $\left\Vert M\right\Vert _{\mathrm{HS}}=
\sqrt{\mathrm{tr}\left(M^{\dagger}M\right)}$
the Hilbert-Schmidt norm of the matrix $M$. Hermitian conjugation
is denoted by $M^{\dagger}$, and transposition by $M^{T}$.

\subsection{Ehrenfest time and spreading of wave packets\label{EhrT}}

It is customary, in this context, to define what is known as the Ehrenfest
time. The latter is a time scale up to which the above approximation
is valid.
We define the Ehrenfest time, denoted $T_{E}\left( \hbar \right)$, as the maximal time  up to which

\begin{itemize}
\item \vspace{\enumskip}the error $\left\Vert \widehat{U}\left(t\right)
\psi_{Z_{0}}^{X_{0}}-e^{\frac{\mathrm{i}}{\hbar}\Theta\left(t\right)}
\psi_{Z_{t}}^{X_{t}}\right\Vert$
remains small, \vspace{\enumskip}
\item the exact state remains localized.\vspace{\enumskip} 
\end{itemize}
The latter ensures that the approximation retains a physical meaning, i.e.,
the above classical approximation makes no sense if the state does not remain 
localized.

The total width of the semiclassically evolved Gau{\ss}ian is
 \begin{eqnarray}
\sigma_{t}\left(X_{0};\,\hbar\right):=\Delta x^{2}\left(\hbar,\, t
\right)+\Delta p^{2}\left(\hbar,\, t\right) & = & \frac{\hbar}{2}
\mathrm{tr}\left(G_{t}\right),\label{eq:semclawidth}
\end{eqnarray}
where we have used the variance
\begin{eqnarray*}
\Delta x^{2} & = & \left\langle e^{\frac{\mathrm{i}}{\hbar}
\Theta\left(t\right)}\psi_{Z_{t}}^{X_{t}},
\, \widehat{x}^{2}e^{\frac{\mathrm{i}}{\hbar}\Theta\left(t\right)}
\psi_{Z_{t}}^{X_{t}}\right\rangle _{L^{2}\left(\mathds{R}^{d}\right)}-
\left\langle e^{\frac{\mathrm{i}}{\hbar}\Theta\left(t\right)}
\psi_{Z_{t}}^{X_{t}},\, \widehat{x}e^{\frac{\mathrm{i}}{\hbar}\Theta
\left(t\right)}\psi_{Z_{t}}^{X_{t}}\right\rangle_{L^{2}
\left(\mathds{R}^{d}\right)}^{2},
\end{eqnarray*}
and similarly for the momentum operator.

From eq. (\ref{eq:width}) one obtains that
\begin{eqnarray*}
\sigma_{t} & \leq & \sigma_{0}\left\Vert S_{t}\right\Vert^{2}_{\mathrm{HS}}.
\end{eqnarray*}
By Theorem \ref{thm:propagation_control_CR}, there will
exist a constant $c_{1}$ such that
\begin{eqnarray*}
\left\Vert\widehat{U}\left(t\right)\psi_{Z_{0}}^{X_{0}}-e^{\frac{\mathrm{i}}{\hbar}
\Theta\left(t\right)}\psi_{Z_{t}}^{X_{t}}\right\Vert & \leq &
c_{1} \sqrt{\hbar} t e^{3\gamma t}.
\end{eqnarray*}
The time scale for which the errors are small is thus algebraic in $\hbar$ if $\gamma=0$. 
In the generic case, the errors remain small for logarithmic times in $\hbar$.

With this result, the errors
\begin{eqnarray}
\Delta\left(t\right) & := & \left\langle \widehat{U}\left(t\right)\varphi_{0},\,
\widehat{s}\left(\widehat{U}\left(t\right)
\varphi_{0}\right)\right\rangle_{L^{2}\left(\mathds{R}^{d}\right)} -\left\langle \widehat{U}_{2}\left(t\right)\varphi_{0},\,\widehat{s}\left(\widehat{U}_{2}
\left(t\right)\varphi_{0}\right)\right\rangle_{L^{2}\left(\mathds{R}^{d}\right)} \label{eq:obserror}
\end{eqnarray}
for propagating observables $\widehat{s}$ can be approximated explicitly. Of particular interest to us is the width
operator $\widehat{s}:=\mathrm{Op}_{\hbar}^{w}\left[\left|Y\right|^{2}\right]$. One can characterize
the times for which the error 
$\Delta\left(t\right)$ is small, e.g., $\Delta\left(t\right)=\mathcal{O}\left(\hbar^{\alpha}\right)$
for some $\alpha>0$. Again, using the Lyapunov inequality (\ref{lyapunovinequality}) one finds that the width 
of the approximate state is bounded by
\begin{eqnarray*}
\sigma_{t}\left(\hbar\right) & \leq & c_{3} e^{2\gamma\left|t\right|},
\quad c_{3}>0,
\end{eqnarray*}
and the error remains $\mathcal{O}\left(\hbar^{\alpha}\right)$ up to times
 of order $\frac{\left|\ln\left(\hbar\right)\right|}{6\gamma}.$ We may thus 
generically state that the Ehrenfest time is
\begin{eqnarray}
T_{E}\left( \hbar \right) & \propto & \frac{\left|\ln\left(\hbar\right)\right|}{6\gamma}.
\end{eqnarray}

In the integrable case, implying  $\gamma=0$, the width grows like the square of the 
Hilbert-Schmidt norm of the flow differential, i.e., at most polynomially in time. The 
error remains small for times up to $\hbar^{-\frac{1}{2}}$. The Ehrenfest time thus 
is algebraic in $\hbar$.

Our aim is to characterize the spreading of the approximate state
for a more specific class of classical motions.

\section{Results}

In addition to the conditions imposed on symbols $H$ above  (see section 
\ref{conditions}), we assume that the Hessian 
$H^{\prime\prime}\left(X_{t}\right)$ is $T-$periodic, i.e., 
\begin{eqnarray}
H^{\prime\prime}\left(X_{t}\right) & = & H^{\prime\prime}\left(X_{t+T}\right),
\quad\forall t\in\mathds{R}.
\label{perhess}
\end{eqnarray}

We will also utilize the following definition.

\begin{definition}
A \textbf{classical revival} at a time $t>0$ is the event that 
the approximate Gau{\ss}ian given above is the 
initial one up to a phase factor, i.e.,
\begin{eqnarray*}
\psi_{Z_{t}}^{X_{t}} & = & e^{\mathrm{i}\alpha_{t}}\psi_{Z_{0}}^{X_{0}}.
\end{eqnarray*}
\end{definition}

\subsection{\label{sec:Floquet-theory}Floquet theory}

According to the Floquet theorem  \cite{F1883}, any  linear differential 
equation with continuous $T-$periodic coefficients has a periodic 
solution of the second type, i.e., a solutions which satisfy
\begin{eqnarray*}
f\left(t+T\right) & = & \upsilon f\left(t\right),\quad
\upsilon\in\mathds{C},\,\forall t\in\mathds{R}.
\end{eqnarray*}
In particular, the linear vector differential equation
\begin{eqnarray*}
\frac{\mathrm{d}}{\mathrm{d}t}f & = & A\left(t\right)f,
\end{eqnarray*}
where $A\left( t\right) $ is continuous and satisfies  $A\left(t+T\right)=A\left(t\right),\,\forall t\in\mathds{R}$, has a
fundamental matrix of the form (see e.g. \cite{CoL55})
\begin{eqnarray*}
F_{t} & = & M^{-1}e^{L t}MU\left(t\right),\quad M\in\mathrm{GL}\left(n,\,
\mathds{C}\right),
\end{eqnarray*} 
where $L$ is a diagonal matrix and $U\left(t\right)$ is a $T-$periodic matrix. 
By definition, a fundamental matrix is a full rank matrix whose columns satisfy 
the differential equation, i.e., the linear combinations of the columns of the 
fundamental matrix span the full space of solutions of the differential equation. 

We will call the elements of $L$ the \textit{Floquet exponents} of the fundamental system.

The result of this section will be summarized in the following way.

\begin{proposition}
\label{pro:periodicwidth} \textbf{If} $H^{\prime\prime}\left(X_{t}\right)$
is periodic \textbf{and} the Floquet exponents of $\,S_{t}$
are purely complex \textbf{then} the width of a Gau{\ss}ian propagated semiclassically
by $\widehat{U}_{2}\left(t\right)$ will remain unchanged at multiples
of the smallest classical period $T$. Furthermore, \textbf{if} the 
classical flow is periodic \textbf{then} classical revivals will occur.
\end{proposition}
\begin{bew}{}
Under the condition (\ref{perhess}), Floquet theory states the 
existence of a fundamental Floquet matrix
for the linear differential eq. (\ref{eq:action}),
\begin{eqnarray*}
S_{t} & = & M^{-1}e^{\Lambda t}M F_{t},\quad 
M\in\mathrm{GL}\left(2d,\,\mathds{C}\right),
\end{eqnarray*}
where $e^{\Lambda t}$ is the diagonal matrix with entries 
$e^{2\pi\lambda_{i}\frac{t}{T}},\,\lambda_{i}\in\mathds{C}$,
$F_{t}$ has minimal period $T$, and we have chosen
\begin{eqnarray*}
F_{0} & = & S_{0} =  \mathds{1}_{2d\times2d}.
\end{eqnarray*}

One directly concludes that
\begin{eqnarray*}
S_{k T} & = & M^{-1}e^{k T\Lambda}M,\quad\forall k\in\mathds{Z}.
\end{eqnarray*}
 for multiples $k$ of the classical period $T$. 

Furthermore, the real fundamental matrix defined by 
$S_{t}\in\mathrm{Sp}\left(2d,\,\mathds{R}\right)$
has a unique polar decomposition, i.e., there exists \cite{Fol89}
an orthogonal matrix $\mathcal{Q}_{t}\in\mathrm{O}\left(2d\right)\cap\mathrm{Sp}\left(2d,\,\mathds{R}\right)$
and a positive definite matrix $\mathcal{P}_{t}\in\mathrm{Sp} \left(2d,\,\mathds{R}\right)$
such that 
\begin{eqnarray*}
S_{t} & = & \mathcal{Q}_{t}\mathcal{P}_{t}.
\end{eqnarray*}
 The width of the approximate squeezed state at time $t$ is 
(see eq. (\ref{eq:semclawidth}))
\begin{eqnarray*}
\sigma_{t}\left(\hbar\right) & = & \mathrm{tr}\left(
\left(S_{t}^{-1}\right)^{T}G_{0}
S_{t}^{-1}\right),
\end{eqnarray*}
and since any symplectic matrix $A$ satisfies \cite{Fol89}
\begin{eqnarray*}
A^{-1} & = & \mathcal{J}A^{T}\mathcal{J}^{-1},
\end{eqnarray*}
we can write by eq. (\ref{eq:width}),
\begin{eqnarray*}
\sigma_{t}\left(\hbar\right) & = & \frac{\hbar}{2}\mathrm{tr}
\left(\mathcal{Q}_{t}\mathcal{P}_{t}\mathcal{J}^{T}
G_{0}\mathcal{JP}_{t}^{T}\mathcal{Q}_{t}^{T}\right)=
\frac{\hbar}{2}\mathrm{tr}
\left(\mathcal{P}_{t}\mathcal{J}^{T}G_{0}\mathcal{J}
\mathcal{P}_{t}^{T}\right),
\end{eqnarray*}
since $\mathcal{Q}_{t}\in\mathrm{O}\left(2d\right)$ .

We are hence confronted with two cases.

\begin{itemize}
\item \vspace{\enumskip} Either $S_{k T}$ is orthogonal, i.e.,
$\mathcal{P}_{k T}=\mathds{1}_{2d\times2d}$; this corresponds to
\begin{eqnarray*}
\sigma_{k T}\left(\hbar\right) & = & \frac{\hbar}{2}\mathrm{tr}
\left(G_{0}\right) = \sigma_{0}\left(\hbar\right),\quad\forall k\in\mathds{Z}.
\end{eqnarray*}
The orthogonality of $S_{k T}$ implies that it has $2d$ singular
values $1$. It is furthermore similar to $e^{k T\Lambda}$
$\left( F_{k T}=\mathds{1}_{2d\times2d}\right)$, which 
hence is unitary so the Floquet
exponents $\frac{2\pi\lambda_{i}}{T}$ are purely complex or zero.
By definition, the Lyapunov exponent of the classical trajectories
is
\begin{eqnarray*}
\gamma & := & \max_{k}\left[ \lim_{t\rightarrow\infty}\sup\left(\frac{\ln\left(
s_{k}\left(X_{t}\right)\right)}{t}\right)\right] ,
\end{eqnarray*}
where $s_{k}\left( t\right)$ are the singular values of the flow differential
$S_{t}$, i.e., the eigenvalues of 
\begin{eqnarray*}
F_{t}^{\dagger}\left(M^{-1}e^{t\Lambda}M\right)^{\dagger}
M^{-1}e^{t\Lambda}M
F_{t} & = & F_{t}^{\dagger}F_{t}.
\end{eqnarray*}
 Noting that $F_{t}$ is bounded (and periodic),
we see directly that if 
\begin{eqnarray*}
\max_{i}\left(\Re\left( \lambda_{i}\right)\right)  & = & 0,
\end{eqnarray*}
the classical motion is linearly stable.\vspace{\enumskip}
\end{itemize}

\begin{itemize}
\item \vspace{\enumskip}$\mathcal{P}_{k T}\neq\mathds{1}_{2d\times2d},$ which 
corresponds to
\begin{eqnarray*}
\sigma_{k T}\left(\hbar\right) & = & \frac{\hbar}{2}\mathrm{tr}
\left(\mathcal{P}_{k T}\mathcal{J}^{T}G_{0}\mathcal{JP}_{k T}^{T}
\right)\\
 & > & \sigma_{0}\left(\hbar\right).\end{eqnarray*}
since $\mathcal{P}_{t}\in\mathrm{Sp}\left(2d,\,\mathds{R}\right)$
is strictly positive definite. This corresponds to the case when $\Re\left(
\lambda_{i}\right)\neq0,$ hence $\max_{i}\left( s_{i}\right)>0$. The classical motion
is hence not linearly stable in this case.\vspace{\enumskip}
\end{itemize}

Furthermore, if the (purely complex) Floquet exponents are
rationally dependent, there exist some multiples 
\begin{eqnarray*}
T_{R} & = & n_{R} T,\quad n_{R}\in\mathds{Z},
\end{eqnarray*}
 of the classical period $T$ such that the orthogonal transformation
at times $T_{R}$ reduces to unity. Indeed, if $n_{R}$ is the smallest common multiple
of the denominators of the sequence $\left\{ \lambda_{i}\right\} $
defined by the Floquet exponents, i.e., 
\begin{eqnarray*}
\lambda_{i} n_{R} & \in & \mathds{N},\,\forall i,
\end{eqnarray*}
 we find
\begin{eqnarray*}
M^{-1}e^{\Lambda T_{R}}M & = & M^{-1}\mathrm{diag}\left\{ e^{2\pi\mathrm{i} 
n_{i}}\right\} M,\quad n_{i}\in\mathds{Z},\\
 & = & \mathds{1}_{2d\times2d}.
\end{eqnarray*}
The semiclassical approximation then is the initial Gau{\ss}ian, if it 
is localized at the initial point $X_{0}$. This is the case if the flow is 
periodic. In this case we have a classical revival. \qed
\end{bew}

\begin{remark}
Note that these revivals are purely classical in the sense that
the conditions only reflect properties of the classical motion
and are hence independent of $\hbar$.
\end{remark}

\begin{remark}
If $S_{k T}$ is orthogonal, the approximate Gau{\ss}ian profile will 
be the initial one at classical periods. It is just rotated and 
translated in phase space.
\end{remark}

\subsection{\label{sec:A-Uniform-bound}A uniform bound}

We can further characterize the approximate state in the case when
the Floquet exponents are purely complex.

\begin{lemma}
\label{lem:Localization} \textbf{If} $H^{\prime\prime}\left(X_{t}\right)$
is $T-$periodic \textbf{and} the Floquet exponents of $S_{t}$
are purely complex \textbf{then} the width $\sigma_{t}\left(\hbar\right)$
of a Gau{\ss}ian state propagated by $\widehat{U}_{2}\left(t\right)$
satisfies the following uniform bound
\begin{eqnarray}
\sigma_{t}\left(\hbar\right) & \leq & 
K\sigma_{0}\left(\hbar\right),\,\forall t\in\mathds{R}.
\label{eq:regularboundedwidth}
\end{eqnarray}
 Furthermore, $K:=e^{\kappa}$, where $\kappa$ is defined by
\begin{eqnarray*}
\kappa & := & 2 T\sup_{t\in\left[0,\, T\right]}\left\Vert \mathcal{J}
H^{\prime\prime}\left(X_{t}\right)\right\Vert _{\mathrm{HS}}.\end{eqnarray*}

\end{lemma}
\begin{bew}{}
We consider eq. (\ref{eq:action}) \[
\left\{ 
\begin{array}{rll}
\frac{\mathrm{d}S_{t}}{\mathrm{d}t} & = & 
\mathcal{J}H^{\prime\prime}\left(X_{t}\right)S_{t}\\
\\S_{0} & = & \mathds{1}.
\end{array}\right.\]
Starting at some initial time $k T,\, k\in\mathds{Z}$, $S_{t+k T}$
satisfies the Gr\"{o}nwall inequality
\begin{eqnarray*}
\left\Vert S_{t+k T}\right\Vert _{\mathrm{HS}} & \leq & 
\left\Vert S_{k T}\right\Vert _{\mathrm{HS}} e^{\kappa},\,\forall t\in\left[0,\, T\right],
\end{eqnarray*}
during a classical period $T$ where 
\begin{eqnarray*}
\kappa & = & 2 T\sup_{t\in\left[0,\, T\right]}\left\Vert 
\mathcal{J}H^{\prime\prime}\left(X_{t}\right)\right\Vert _{\mathrm{HS}}.
\end{eqnarray*}
Furthermore,
 \begin{eqnarray*}
\sigma_{t}\left(\hbar\right) & = & \frac{\hbar}{2}\mathrm{tr}
\left(S_{t}\mathcal{J}^{T}G_{0}\mathcal{J}
S_{t}^{T}\right)\\
 & = & \frac{\hbar}{2}\mathrm{tr}\left(G_{0}
S_{t}^{T}S_{t}\right)\leq\frac{\hbar}{2}\left|
\mathrm{tr}\left(G_{0}\right)\right|\left\Vert 
S_{t}\right\Vert _{\mathrm{HS}}^{2}.
\end{eqnarray*}
Hence
\begin{eqnarray}
\sigma_{t}\left(\hbar\right) & \leq & \frac{\hbar}{2}\left|
\mathrm{tr}\left(\mathbf{g}_{0}\right)\right| e^{\kappa},\label{eq:boundbetweenpewriods}
\end{eqnarray}
if $S_{t}$ has purely complex Floquet exponents since in
this case $\sigma_{0}\left(\hbar\right)=\sigma_{k T}\left(\hbar\right),$
according to proposition \ref{pro:periodicwidth}. 

We conclude with this choice that
\begin{eqnarray*}
\sigma_{t}\left(\hbar\right) & \leq & K\sigma_{0}\left(\hbar\right)\end{eqnarray*}
uniformly in time. \qed
\end{bew}
If $S_{t}$ is not orthogonal at the classical periods $T$,
the latter bound takes the form
\begin{eqnarray*}
\sigma_{t}\left(\hbar\right) & \leq & K\sigma_{0}\left(\hbar\right) 
e^{2\max_{i}\left(\Re\left(l_{i}\right)\right)\left|t\right|}
\end{eqnarray*}
where $l_{i}:=\frac{2\pi\lambda_{i}}{T}$ are the corresponding
Floquet exponents of $S_{t}$ which is nothing else than a
Lyapunov inequality where $\max_{i}\left(\Re\left(l_{i}\right)\right)$
plays the role of the Lyapunov exponent. Using the fact that
what we are approximating is the Hilbert-Schmidt norm of the flow
differential one can furthermore state the existence of a constant
$c_{4}>0$ such that 
\begin{eqnarray*}
\frac{e^{2\max_{i}\left(\Re\left(l_{i}\right)\right)\left|t\right|}}{c_{4}} & \leq\sigma_{t}
\left(\hbar\right)\leq & c_{4} e^{2\max_{i}\left(\Re\left(l_{i}\right)\right)\left|t\right|}.
\end{eqnarray*}
which determines the asymptotic behavior of the width in that case.

\begin{remark}
We note that the equality in eq. (\ref{eq:regularboundedwidth}) is
reached if the classical period is null, i.e., if the Hamiltonian
has a constant Hessian $H^{\prime\prime}\left(X_{t}\right)$.
This is satisfied by the harmonic oscillator. The flow differential 
$S_{t}$ is then an orthogonal matrix for all
times. It is well known that the propagation is dispersion-less in
this case.
\end{remark}

\section{\label{sec:Discussion}Discussion}

Theorem \ref{thm:propagation_control_CR} allows us to trace back
the properties of the approximation to the exact state
$\mathcal{U}\left(t\right)\psi_{Z_{0}}^{X_{0}}$. According to the discussion in
\cite{CR97} (see also section \ref{EhrT}), this can be done with an error 
of order $\mathcal{O}\left(\hbar^{\alpha}\right),\,\alpha>0$ up to 
times $\hbar^{-\frac{1}{2}}$ if the Floquet exponents have zero real 
part (stable classical dynamics) and up to times $\frac{\left|\ln\left(
\hbar\right)\right|}{6 \nu},\,\nu=\max_{i}\Re\left(l_{i}\right)$
otherwise (unstable classical dynamics) as 
$\hbar \searrow 0$ and $t \nearrow \infty$.

In the stable case, the approximate state remains localized ad infinitum
 if this state remains localized between classical periods since
the width of such a state is the same at classical periods. We wish to 
stress that we have defined the classical period as the minimal period 
of the Hessian  $H^{\prime\prime}\left(X_{t}\right)$ which is not necessarily 
the period of the classical flow $\Phi_{H}^{t}$. An example is classical motion 
in a one dimensional periodic potential $V_{\Gamma}$ with energies $E$ such that
\begin{eqnarray*}
E & > & \sup_{x\in\mathds{R}}
\left(V_{\Gamma}\left(x\right)\right).
\end{eqnarray*}
In such a case $H^{\prime\prime}\left( X_{t}\right)$ is periodic although the flow
$\Phi_{H}^{t}\left( X_{0}\right) $ is not. Furthermore, the approximation remains
localized between classical periods if 
\begin{eqnarray*}
\kappa &  = & 2T
\sup_{t\in\left[0,\,T\right]}\left\Vert \mathcal{J}H^{\prime\prime}\left(X_{t}\right) 
\right\Vert_\mathrm{HS}, 
\end{eqnarray*}
the exponent in eq. (\ref{eq:boundbetweenpewriods}), 
is small enough. We can hence state in that case, up to small errors 
$\mathcal{O}\left(\hbar^{\alpha}\right)$, localization of the exact 
propagation up to the Ehrenfest time $T_{E}\propto
\hbar^{-\frac{1}{2}}$. Note that the situation includes
unbounded motion.

If the flow differential $S_{t}$ is periodic, it is clear that
the shape will be the initial one at classical periods up to a small 
error. The resonance condition allows to state that this will also
occur at some time if the flow differential is merely periodic of the
second type and if the Floquet exponents are rationally 
dependent and purely imaginary. We may state this since the exact state remains 
localized and that our asymptotic considerations are valid as $\hbar 
\searrow 0$ and  $t \nearrow \infty$. The property extends also to the 
general stable case, i.e., in the case of rationally independent
Floquet exponents. Recall that the quadratic form $Z_{t}$ is given by
linear fractional transformation. In particular,
\begin{eqnarray*}
Z_{kT} & = & \left(A_{kT}Z_{0}+B_{kT}\right)\left(C_{kT}Z_{0}+D_{kT}\right)^{-1}
\end{eqnarray*}
where we have used the notation of section \ref{sec:A semiclassical propagation theorem}.
Since $S_{kT}\sim e^{\Lambda kT}$ which is unitary with
eigenvalues $e^{\mathrm{i}\ell_{i}\frac{2\pi}{T}}=e^{\lambda_{i}\frac{2\pi}{T}}$ where
$\ell_{i}\in\mathds{R}$, there exist $k-$independent matrices 
$a_{i}\in\mathds{C}^{d\times d}$ such that $Z_{kT}$ can be viewed as the image of the vector 
$\ell\in\mathds{T}^{d}$ with entries $\ell_{i}$ under the continuous map
\begin{eqnarray*}
\mathsf{e} & : & \mathds{T}^{d}\rightarrow\Sigma_{d}\\
 &  & \omega\mapsto\sum_{i=0}^{d}a_{i}e^{2\pi\mathrm{i}\omega_{i}}.
\end{eqnarray*}
Furthermore, the endomorphism
\begin{eqnarray*}
\tau & : & \mathds{T}^{d}\rightarrow\mathds{T}^{d}\\
 &  & L \mapsto L+\ell,
\end{eqnarray*}
where $\left( L_{i}+\ell_{i}\right)$ is defined modulo 1, is known to be ergodic with respect to Lebesgue measure on the $d-$torus since the frequencies $\ell_{i}$
are rationally independent. In this notation we have $Z_{0}=\mathsf{e}\left(0\right)=
\mathsf{e}\left(L\left( 0\right) \right)$.
From the ergodicity of $\tau$ on $\mathds{T}^{d}$ we may state that
for every $\varepsilon>0$ there exist $K\in\mathds{Z}$ such
that $\left\Vert \tau^{K}\left(L\left( 0\right) \right)-L\left( 0\right) \right\Vert <\varepsilon$.
It follows that for every $\epsilon>0$ there exist
some $n\in\mathds{Z}$ such that $\left\Vert \mathsf{e}\circ\tau^{n}\left(0\right)-\mathsf{e}\left(0\right)\right\Vert <\epsilon$ implying
\begin{eqnarray*}
\forall \epsilon > 0 , \, \exists n\in\mathds{Z},\quad\left\Vert Z_{n T}-Z_{0}\right\Vert < \epsilon, 
\end{eqnarray*}
i.e., $Z_{kT}$ is quasi-periodic.
One concludes that the initial shape of the approximated state will up to some small 
error reoccur at some multiple of the classical period if the Floquet exponents of the 
flow differential are purely imaginary. We have again no {\`a} priori reasons to exclude unbounded motion.

In the case of a periodic flow, we have localization at the initial point and the 
initial profile (up to some small error) at classical periods. For 
bounded classical motion, those recurrences define revivals at periods 
of the classical flow that have already been described \cite{APe89} 
in the past.

We summarize our findings in the following theorem.

\begin{theorem}
Under the following assumptions 

\begin{enumerate}
\item $H\left(X\right)$ satisfies the conditions of section \ref{conditions}
\vspace{\enumskip}
\item $H^{\prime\prime}\left(X_{t}\right)$ is $T-$periodic and
$\nu$ is the maximal real part of the Floquet exponents of the flow
differential, solution of eq. (\ref{eq:action}),\vspace{\enumskip}
\vspace{\enumskip}
\end{enumerate}
and with $\kappa:=2 T\sup_{t\in\left[0,\, T\right]}\left\Vert \mathcal{J}H^{\prime\prime}\left(X_{t}\right)\right\Vert _{\mathrm{HS}}$ where $K:=e^{\kappa},$
\vspace{\enumskip}

\noindent we can make the following statements up to $\mathcal{O}\left(\hbar^{\alpha}\right),\,\alpha>0$
as $\hbar\searrow0$.

\textbf{If} $K$ is small enough \textbf{then} the approximation described by theorem \ref{thm:propagation_control_CR}
will hold up to times $\frac{\left|\ln\left(\hbar\right)\right|}{6\nu}$
and the approximate width of the state will behave like $e^{2\nu\left|t\right|}$.

\textbf{In particular,} \textbf{if} the Floquet exponents are purely
complex or zero \textbf{then} the semiclassical propagation described
in theorem \ref{thm:propagation_control_CR} will hold to times $\hbar^{-\frac{1}{2}}$.
The width of the approximated state will be bounded and recurrences will, up to small errors, 
take place at a multiple of $T$.
\end{theorem}
\begin{acknowledgement}
We would particularly like to thank Jens Bolte for valuable comments
and discussions.
\end{acknowledgement}
\bibliographystyle{unsrt}
\bibliography{/home/estraeng/Desktop/sp_prop/bib_jes}

\begin{thebibliography}{10}

\bibitem{CR97}
M.~Combescure and D.~Robert.
\newblock Semiclassical spreading of quantum wave packets and applications near
  unstable fixed points of the classical flow.
\newblock {\em Asymptot. Anal.}, 14(4):377--404, 1997.

\bibitem{PaUr93}
Thierry Paul and Alejandro Uribe.
\newblock A construction of quasi-modes using coherent states.
\newblock {\em Ann. Inst. H. Poincar\'e Phys. Th\'eor.}, 59(4):357--381, 1993.

\bibitem{Har04}
T.~Hartmann, F.~Keck, H.~J. Korsch, and S.~Mossmann.
\newblock Dynamics of bloch oscillations.
\newblock {\em New J. Phys.}, 6(2), 2004.

\bibitem{Witt04}
D.~Witthaut, F.~Keck, H.~J. Korsch, and S.~Mossmann.
\newblock Bloch oscillations in two-dimensional lattices.
\newblock {\em New J. Phys.}, 6(41), 2004.

\bibitem{Blo28}
F.~Bloch.
\newblock \"{U}ber die quantenmechanik der elektronen in kristallgitter.
\newblock {\em Zeitschrift f\"{u}r Physik}.

\bibitem{Ha81}
George~A. Hagedorn.
\newblock Semiclassical quantum mechanics. {III}. {T}he large order asymptotics
  and more general states.
\newblock {\em Ann. Physics}, 135(1):58--70, 1981.

\bibitem{Ha85}
George~A. Hagedorn.
\newblock Semiclassical quantum mechanics. {IV}. {L}arge order asymptotics and
  more general states in more than one dimension.
\newblock {\em Ann. Inst. H. Poincar\'e Phys. Th\'eor.}, 42(4):363--374, 1985.

\bibitem{Fol89}
Gerald~B. Folland.
\newblock {\em Harmonic analysis in phase space}, volume 122 of {\em Annals of
  Mathematics Studies}.
\newblock Princeton University Press, Princeton, NJ, 1989.

\bibitem{F1883}
G.~Floquet.
\newblock Sur les \'equations diff\'erentielles lin\'eaires \`a coefficients
  p\'eriodiques.
\newblock {\em Ann. Sci. \'Ecole Norm. Sup. (2)}, 12:47--88, 1883.

\bibitem{CoL55}
Earl~A. Coddington and Norman Levinson.
\newblock {\em Theory of ordinary differential equations}.
\newblock McGraw-Hill Book Company, Inc., New York-Toronto-London, 1955.

\bibitem{APe89}
I.Sh. Averbuck and N.F. Perelman.
\newblock Fractional revivals: universality in the long-term evolution of
  quantum wave packets beyond the correspondence principle dynamics.
\newblock {\em Physics Letters A}, 139:449--453, 1989.

\end{thebibliography}
\end{document}